\documentclass[thmsa,11pt]{article}
\usepackage{sw20lart}
\usepackage{amsmath}
\usepackage{graphicx}
\usepackage{geometry}
\usepackage{amsfonts}
\usepackage{amssymb}
\begin{document}

\author{\textbf{Scott M. Hitchcock}\\National Superconducting Cyclotron Laboratory (NSCL)\\Michigan State University, East Lansing, MI 48824-1321\\E-mail: hitchcock@nscl.msu.edu}
\title{\textbf{Feynman Clocks, Causal Networks, and the Origin of Hierarchical
'Arrows of Time' in Complex Systems from the Big Bang to the Brain'}\\{\small An Invited Talk Given at The Institute For High Energy Physics,
Protvino, Russia.}\textbf{\ }}
\date{Presented: Wednesday, June 21, 2000}
\maketitle
\begin{abstract}
A theory of \textbf{time} as the '\textbf{information'} created in the
\emph{irreversible decay process of excited or unstable states} is
proposed\emph{.} Using new tools such as \textbf{Feynman Clocks (FCs)
\cite{hitchcock}, \cite{hitchcock2}, }\cite{Peres}\textbf{,} \textbf{Feynman
Detectors (FDs),} \textbf{Collective Excitation Networks (CENs)},
\textbf{Sequential Excitation Networks (SENs), Plateaus of Complexity (POCs),
Causal Networks, }and\textbf{\ Quantum Computation Methods }\cite{asher},
\cite{eiqc}, \cite{phys229}\textbf{\ }previously \emph{separate}
'\textbf{arrows of time}' describing change in complex systems ranging from
the Big Bang to the emergence of consciousness in the Brain are 'unified'. The
'\textbf{direction}' and '\textbf{dimension}' of time are created from clock
ordered sets of real number 'labels' coupled to signal induced states in
detectors and memory registers. The '\emph{Problem of Time}' can be 'solved'
using the fundamental \emph{irreversible} \textbf{Quantum Arrow of Time (QAT)}
and \emph{reversible} \textbf{Classical Arrows of Time (CATs) }to 'map'
information flow in causal networks\textbf{.} A pair of communicating
electronic Feynman Clock/Detector units were built and used to demonstrate the
basic principles of this new theory of 'time'.
\end{abstract}

\section{Introduction}

\begin{quotation}
\emph{Should we be prepared to see some day a new structure for the
foundations of physics that does away with time?...Yes, because ''time'' is in
trouble.-John Wheeler }\cite{davies}\emph{.}
\end{quotation}

It has been suggested by Julian Barbour that 'time' does not exist
\cite{timeless} thus adding one more complication to the '\emph{Problem of
Time}'. It is the position of this author that 'time' in fact 'exists'. 'Time'
is a different 'property' of evolving systems than has been previously
assumed. This problem was explored last year at the XXII International
Workshop on High Energy Physics and Field Theory by \emph{C. Marchal}
\cite{marchal}. I will take a different approach that will in principle
provide a basis for the 'statistical' or 'ensemble'\emph{\ arrows of time} of
Poincar\'{e} as the result of $n$-body causal networks of fundamental quantum
systems called \textbf{Feynman Clocks }or\textbf{\ FCs}.

The 'Problem of Time' is two fold. First there is the problem of \emph{what}
is 'time'? Second is the problem of the \emph{apparent irreversibility} in a
macroscopic world built on the \emph{apparent reversibility} of the
microscopic world. The first question can be answered by identifying
'\textbf{time}' as a form of \textbf{information}. The second question can be
answered by identification of an \emph{irreversible} \textbf{Quantum Arrow of
Time (QAT) }and \emph{reversible} \textbf{Classical Arrows of Time (CATs).
}The QAT and CATs are\textbf{\ pointers} mapping information flow between
quantum systems and their environments in hierarchical causal networks. QATs
are pointers generated by transformations between quantum states in a finite
'lifetime'. CATs are pointers whose 'endpoints' are the 'time labels' for
events. The causal separation between these events is the classical 'time
difference' interval or the \emph{magnitude} of the CATs.

The \emph{correspondence} between the various separate biological,
cosmological, psychological, radiative, and thermodynamic '\emph{arrows of
time'} \cite{prig}, \cite{zeh}, \cite{cove}, \cite{schulman1}, \cite{rirnr} is
achieved with \emph{causal networks}. Hierarchical \textbf{Plateaus of
Complexity (POCs) }emerge at various size scales (e.g. atoms, molecules, and
cells)\textbf{. }Feynman Clocks (a general form of a 'quantum clock'
\cite{Peres})\textbf{, Collective Excitation Networks or CENs , }%
and\textbf{\ Sequential Excitation Networks or SENs} are the '\textbf{nodes}'
and '\textbf{gates}' used to build causal networks.

We begin with the idea that \textbf{time is a number created by the processing
of 'energy of reconfiguration' information (}with dimensions of 'energy')
carried by '\textbf{signals}'. The signal information represents the endpoint
of a geometric and energy configuration change of the matter in a source
system.\ Understanding the internal structure of a system by interpretation of
it's signals requires more than one \emph{bit, byte} or \textbf{qubit} of
information in order to identify the '\emph{geometry}' of the matter. A good
example of this is the \emph{deduction} of the Bohr energy level model for the
Hydrogen atom based on the emission line spectra of photon 'signals'. These
signals are the key to mapping the discrete quantized energy levels of the
electron and proton system. It is clear that more than one spectral line is
necessary to understand this system as a \emph{whole}.

In the case of the QAT, scalar division of Planck's constant by the energy of
reconfiguration creates a \emph{real} number representing the
'\textbf{lifetime}' of the decay process (with dimensions of 'time' or
seconds) in a \textbf{Feynman Clock}. The \emph{information} coupled to the
\textbf{signals} emitted by \textbf{Feynman Clocks (FCs)} is transferred to
the FC's 'environment' (e.g. the vacuum) through which it propagates. The
trajectory of the signal may end at a\textbf{\ Feynman Detector (FD)} where it
is converted into an 'excited' state by 'absorption'. The FD is the signal
absorption mode of the FC and unless otherwise indicated 'FC' will be used to
represent these two modes of a \emph{single system}. The conversion of this
state information is called \textbf{Signal} or \textbf{State Mapping}.

This mapping process can be either a 'quantum' or classical computation in a
\emph{connected} set of logic 'gates' or 'nodes' forming a causal network.
This process couples or \textbf{entangles} a '\textbf{number state}' or
'\textbf{time label}' with a '\textbf{standard clock signal' (pulse)
}\emph{and} the \textbf{signal induced state} in a detector or memory
register. This new composite \textbf{coherent 'entangled' }or\textbf{\ triplet
state} \emph{encodes} the '\textbf{event time}' of classical and relativistic
mechanics. The conversion of this entangled state into a real number
representation of the 'event time' results from a selective decoupling or
disentangling measurement on the 'number state' with an attendant loss of
information about the two other states.

In order to create a conventional 'time' for an 'event' the label part of the
triplet state must be measured. The 'signal state' and the 'standard clock
pulse' information is lost as part of the \emph{measurement induced}
\textbf{decoherence }\cite{zeh}, \cite{dacwqt}, \cite{zurek} of the triplet
state. This loss is indicated by an increase in the 'entropy' of the
environment or the gate in which the processing of the triplet state occurred.
The resulting numerical label is a singlet state that can be used to
\textbf{build} the conventional 'dimension' and 'direction' of 'time' in
space-time using ordered sets of labeled singlet states representing the
causal order of detected events with respect to a standard or internal clock.

The FC model may be applied to quantum cosmology and the Big Bang origin of
the universe up to a point at which the vacuum appears as an 'environment'
decoupled from the rest of the matter and energy. At this point, 'freezing
out' of the vacuum decouples the FC-Universe QAT into 'local' QATs associated
with all the various quantum subsystems distributed throughout the universe.
Classical CATs emerge from the interactions of information provided by the
environment to these quantum systems. The separation of these quantum systems
in space is necessary for 'classical' behaviors to emerge. This separation
also gives rise to hierarchical networks of matter and energy formed, in part,
by gravitational clustering and \emph{information enrichment} in local regions
of space such as the life-forms on our planet's surface.

The expansion of space leads to classical 'entropy' and thermodynamic
considerations defined by separation and unavailability of signals to reset or
create new unstable states of quantum systems. The Big Bang starts as a
quantum system but branches into locally complex quantum and classical systems
spread through out space. The evolution of these systems into hierarchical
networks of ever increasing complexity leads to local high information density
structures of matter such as the brain. The complex states supported by
biological systems can create information 'spikes' in space in contrast with
extreme gravitational information 'spikes' in which information is localized
but trapped in objects like black holes. A balance of gravity and information
flow through a 'system' such as the Earth's ecosphere is necessary for the
emergence of 'consciousness' from hierarchical causal networks of elements
formed in a Big Bang Feynman Clock model and evolved in the complex neural
network systems of the Brain.

\section{The Quantum Arrow of Time (QAT)}

The\textbf{\ Quantum Arrow of Time }or\textbf{\ QAT }is defined by the
\emph{irreversible} 'decay of a discrete unstable state resonantly coupled to
a continuum of final states' \cite{Diu}. The unstable state of a system
originates in the fundamental geometric asymmetry of non-equilibrium spatial
configurations of particles with 'charge', 'spin', and other quantum
properties. The instability is characterized by \emph{finite}
'\textbf{lifetimes}' of these states. The irreversible QAT is observed in the
decay of radioactive nuclei, excited electronic states of atoms through
'autoionization' (emitting a 'free' electron) or photon emission for example.
The decay of these quantum systems can be described with
'\textbf{time-independent}' perturbation theory;

\begin{quotation}
''For example, a system, initially in a discrete state, can split, under the
effect of an internal coupling (described, consequently, by a time-independent
Hamiltonian $W$), into two distinct parts whose energies (kinetic in the case
of material particles and electromagnetic in the case of photons) can have,
theoretically, any value; this gives the set of final states a continuous
nature...We can also cite the \emph{spontaneous emission }of a photon by an
excited atomic (or nuclear) state: the interaction of the atom with the
quantized electromagnetic field couples the discrete initial state (the
excited atom in the absence of photons) with a continuum of final states (the
atom in a lower state in the presence of a photon of arbitrary direction,
polarization and energy).''\cite{Diu}
\end{quotation}

\ These decay modes are not restricted to atoms and nuclei. We will see that
these \emph{Feynman clocks} are \emph{time-independent irreversible systems}
that can be created in space by \emph{apparently} 'time' reversible particle
collisions. \emph{The key to the irreversibility in quantum systems is the
creation of an \textbf{unstable} configuration of matter and energy in space}.
This is the '\emph{first cause}' for decay. \emph{Instability} is a measure of
the \emph{geometric asymmetry} of the mass-energy distribution of the
'components' as they are driven to 'more stable' configurations by the
fundamental interactions (forces) between each other.

We will see that apparent 'time' reversibility and irreversibility are
compatible and necessary aspects of quantum systems. The '\emph{Program of
Decoherence}'\cite{zeh}, the entanglement of quantum states, and the emergence
and decay of novel collective excitations provide tools \cite{naqw},
\cite{qdae}, \cite{hologram} for understanding the common roots of all arrows
of time for unstable configurations of hierarchically scaled clusters of
matter in an evolving universe. This scaling leads to the emergence of
collective 'classical' or macroscopic aspects of reality. The definitions used
in this new approach to the problem of time are:

\begin{definition}
'\textbf{Time}' is a \textbf{form} of '\textbf{information'}.
\end{definition}

\begin{definition}
The \textbf{irreversible} \textbf{Quantum Arrow of Time (QAT)} is always
defined as a \textbf{pointer} from the \textbf{unstable state} to the
\textbf{decay state} for any \textbf{unstable quantum system}.
\end{definition}

\begin{definition}
\textbf{Signals} created by \textbf{reconfigurations} of \textbf{unstable
states of quantum systems (Feynman Clocks or FCs) carry 'Information' to other
systems (Feynman Detector mode of a FC) in Causal Networks}. Signals can be
used to 'reset' an unstable state of a system which will 'decay' irreversibly.
\end{definition}

\begin{definition}
The '\textbf{energy' of reconfiguration} represents the \textbf{information
content }of the transition from the initial to final wave-functions via the
reconfiguration operator or Hamiltonian. The energy of reconfiguration is
converted into the '\textbf{lifetime}' of the \textbf{unstable state }by
division into\textbf{\ Planck's Constant}.
\end{definition}

\begin{definition}
The signal induced state \textbf{information} in a FD or memory register is
converted or 'computed' into a \textbf{time number }or \textbf{time label} by
\textbf{signal} or \textbf{state} '\textbf{mapping}'. The signal induced state
is mapped to a '\textbf{standard clock pulse}' and a concurrent
'\textbf{number state}' creating a new \textbf{composite} or \textbf{entangled
coherent }'\textbf{triplet}' state. This triplet state can then be converted
into a single real '\textbf{event time' number} by \textbf{disentanglement}
due to a \textbf{measurement} or \textbf{decoherence} process. This real event
time number is the time label used to find the \textbf{classical time
difference} between two time labeled events.
\end{definition}

\begin{definition}
The \textbf{ordered set of real numbers }extracted from triplet states created
by signal events can be used to construct the '\textbf{dimension}' and
'\textbf{direction}' of the \textbf{conventional time} used in classical and
relativistic physics by mapping of the set onto the geometric real number line.
\end{definition}

\section{Signals}

A \textbf{signal} is any '\textbf{system}' (e.g. photon plus vacuum) that
conveys \emph{state} \textbf{information} from one system (e.g. FC) to another
(e.g. FD). The creation of a detector state from a signal state is the end
process of information transfer originating in a spatially distinct FC. The
state information transfer causes the reconfiguration of the detection system
resulting in an unstable state of 'excess' information. This information can
then be converted into the 'time' of detection of the signal through the
signal mapping process.

The classical \emph{transit time} of a 'signal' is given by the 'distance'
travelled by the signal divided by the sinal velocity. This transit time can
be treated as the '\emph{lifetime}' of a\textbf{\ pseudo-FC} or \textbf{pFC}.
The '\emph{excited state}' of this pseudo-FC is the point of creation of a
signal. The decay of this state creates a signal that travels through space to
the '\emph{ground state'} of the vacuum at the point of detection in the FD.
The pseudo-lifetime of this spatially extended quantum system is:%

\begin{equation}
\tau_{signal}=\frac{d_{FC\Rightarrow FD}}{v_{FC\Rightarrow FD}}=\frac{\left|
d_{FD}-d_{FC}\right|  }{v_{FC\Rightarrow FD}}=\mathbf{\tau}_{\mathbf{pFC}%
}\equiv\frac{\hbar}{\Gamma_{FC}}%
\end{equation}

where $\Gamma$\ is the \textbf{'energy of reconfiguration' information }with
dimensions of 'energy' (see section 4 below) given \cite{Veltman}\ by (see the
next section below for details):%

\begin{equation}
\Gamma_{FC}=%
{\displaystyle\idotsint\limits_{q_{n}\cdots q_{1}}}
\left[  \tfrac{V^{n+1}}{\left(  2\pi\right)  ^{3n+4}}\mathbf{P\cdot}\left|
\mathbf{M}_{I}\right|  ^{2}\delta_{4}\left(
{\textstyle\sum\limits_{i=1}^{m}}
\mathbf{p}_{i}-%
{\textstyle\sum\limits_{j=1}^{n}}
\mathbf{q}_{j}\right)  \right]  dq_{1}dq_{2}\cdots dq_{n}%
\end{equation}

This may appear to be rather artificial but by treating the composite system
composed of; the point of emission, the inter-nodal space (e.g. vacuum), and
the point of detection as a \emph{single quantum system}, the 'entanglement'
of separated states of remote detectors can be viewed as the 'collective
excitation' extended over space. \emph{The entangled state of this system
before 'measurement' or decoherence is a collective excitation.} CEs in such
systems may provide explanations for faster than light information flow
between states undergoing disentanglement. This will be useful later when
considering synchronized states of distant gates in quantum computers, CE
states in neural networks and \emph{the universe as a quantum computer
}\cite{lloyd}.

\section{Feynman Clocks (FCs)}

Feynman diagrams are the source of \textbf{Feynman Clocks (FCs)} created by
transformation of the 'time' component (dimension) of the incoming and
outgoing signals into the state information content of those signals. The
interaction (collision or scattering) of the incoming signals can create a
Feynman clock where there was no pre-existing matter before. If there was
matter forming a 'target' for these signals then it acts like the
\textbf{Feynman Detector (FD) mode} of a FC. The target 'detects' the signals
creating new states of the composite system. If this system is \emph{unstable}%
, then the Feynman detector mode of the target has become a Feynman clock. In
general, the incoming particle 'signals' create a clock where there was no
clock before. FCs may be '\emph{open}' or '\emph{closed}' in relation to the
incoming and outgoing signal trajectories. Open FCs are characterized by
spatially 'random' or quasi-wired trajectories for the incoming and outgoing
signals such as in the case for nuclei in stellar environments. Closed FCs are
characterized by 'wired' or 'fixed' incoming and outgoing signal trajectories
in spatially linked networks such as those in optical and electrical circuits.
FCs can be dynamically rewired as an evolutionary process where they can
become closed or open components in causal networks. These are the 'gates' of
quantum computers or microtubule 'processors' in neurons.

For incoming signals whose total momentum is;%

\begin{equation}
\mathbf{p}_{0}=%
{\textstyle\sum\limits_{i=1}^{m}}
\mathbf{p}_{i}%
\end{equation}

resulting in the creation of outgoing signals whose total momentum is;%

\begin{equation}
\mathbf{q}_{0}=%
{\textstyle\sum\limits_{j=1}^{n}}
\mathbf{q}_{j}%
\end{equation}

A 'transient' local quantum clock system is created through reconfigurations
of the matter and energy in the signals via the strong, electromagnetic, weak,
and gravitational fundamental interactions (indexed by $I=s,em,w,g$
respectively). The \textbf{net }Feynman clock 'lifetime' from the system state
created by the interacting incoming signals (FD mode) through the 'decay'
process (internal 'decoherence' mode collective excitation state decay) to the
state in which the outgoing decoupled signals are emitted (FC mode) is given by;%

\begin{align}
\mathbf{\tau}_{FC_{net}}  &  =\tfrac{\hbar}{%
{\displaystyle\idotsint\limits_{q_{n}\cdots q_{1}}}
\left[  \tfrac{V^{n+1}}{\left(  2\pi\right)  ^{3n+4}}\mathbf{P\cdot}\left|
\mathbf{M}_{I}\right|  ^{2}\delta_{4}\left(  \mathbf{p}_{0}-\mathbf{q}%
_{0}\right)  \right]  dq_{1}dq_{2}\cdots dq_{n}}\\
&  =\tfrac{\hbar}{%
{\displaystyle\idotsint\limits_{q_{n}\cdots q_{1}}}
\left[  \tfrac{V^{n+1}}{\left(  2\pi\right)  ^{3n+4}}\mathbf{P\cdot}\left|
\mathbf{M}_{I}\right|  ^{2}\delta_{4}\left(
{\textstyle\sum\limits_{i=1}^{m}}
\mathbf{p}_{i}-%
{\textstyle\sum\limits_{j=1}^{n}}
\mathbf{q}_{j}\right)  \right]  dq_{1}dq_{2}\cdots dq_{n}}%
\end{align}

If there is no reconfiguration of the incoming signals and target (if any) in
this region of space, then a clock has not been 'created' and the reduced
fundamental interaction matrix element $M_{I}$ \cite{Veltman} is zero.

The above equations for the Feynman diagram method for FD/FC 'lifetimes'
represent the creation of 'lifetime' information from a scattering process
that in general is very difficult to compute for complex systems. The idea
here is that a 'collective excitation system' is created by the incoming
signals leading to an irreversible decay with the production of outgoing
signals. The transformation of the incoming signals by collisional
'processing' in a target 'gate' creates new information in the form of the
novel emergent signal states.

Feynman clocks are quantum clocks with multiple inputs and outputs. These
'time-independent' quantum systems are modeled from techniques used in Feynman
Diagrams. These FCs are the basic quantum systems responsible for the
evolution and change in hierarchical structures of matter and the causal
networks they form. For a \textbf{general Feynman Clock }the
intrinsic\textbf{\ lifetime, }$\mathbf{\tau}_{FC}$\textbf{,} is:%

\begin{equation}
\mathbf{\tau}_{FC}=\frac{\hbar}{\mathbf{\Gamma}_{I}}%
\end{equation}

The set of all available information about a system is needed to construct the
wavefunctions and the transformation operator for the FC. The need for
\emph{complete information} creates a problem for the 'observer' since
determination of the FC lifetime as a '\emph{result}' of a theoretical model
is difficult to realistically model except for special cases where the
specific properties of a system in a given set of conditions is known. This is
why the 'observer' or observing system has such an important role in
characterizing the change in causal networks.

\section{Collective Excitations and Entangled States}

The key to single collective behaviors of many quantum systems coordinated by
signals and then acting as 'classical' objects are \textbf{collective}
\textbf{excitations (CEs)} of \textbf{quasiparticles }(also called 'elementary
excitations') \cite{Mattuck}, \cite{zago}, \cite{heyde}, and \textbf{entangled
states}. \textbf{Phonons}, \textbf{excitons}, and \textbf{plasmons} are
examples of CEs that exhibit mesoscopic system behaviors but are still quantum
phenomena \cite{zim}, \cite{ssp}. Entangled states are observed in correlated
behaviors of photons in non-locality experiments in which their initial
coupling at the source of their creation remains even though their physical
space separation is so large that 'signals' travelling at the speed of light
are too slow to account for the communication of state information between the
distant photons when one or both of them are 'measured'\cite{qcent},
\cite{genFC}.

One of the fundamental questions about collective excitations is what are the
maximum distances between CEN components that will still support collective
behavior? The emergence of 'quantum' excitations in mesoscopic
(quasi-classical) and macroscopic CEN systems requires resonant communication
or '\emph{synchronized}' \emph{entanglement} of the states of all the relevant
components. \textbf{Entanglement} of the quantum states of two or more
components provides a composite complex state that can represent a collective
excitation of two 'isolated' but 'historically' coupled signals.

The 'entanglement of states' of nodes in causal networks that have space-like
separations is an essential aspect of behaviors of large numbers of coupled
systems acting as a single collective system with collective excitations that
define the system arrow of time. The entanglement of states is represented by
the 'Direct' or 'Tensor' Products of the individual states of the components
forming the Composite Quantum system. For the case of a composite system,
$\left|  T_{j}\right\rangle $, composed of three entangled subsystems $\left|
t_{j}\right\rangle $, $\left|  q\right\rangle $, and $\left|  \Psi
_{j}\right\rangle $ the '\emph{direct'} or \emph{tensor product} is:%

\begin{equation}
\left|  T_{j}\right\rangle {\LARGE \equiv}\left|  t_{j}\right\rangle
{\LARGE \otimes}\left|  q\right\rangle {\LARGE \otimes}\left|  \Psi
_{j}\right\rangle
\end{equation}

For the case of the\emph{\ triplet state} the \textbf{pulse counter label
state }$\left|  t_{j}\right\rangle $ is a \emph{singlet} state (e.g. 'time
label') such that $t_{j}\in\Re$, where $\Re$ is the set of Real Numbers.

The \textbf{standard clock pulse} or '\textbf{qubit'} \textbf{state} $\left|
q\right\rangle $\ is a \emph{superposition} of two standard clock pulse
states, $\left|  0\right\rangle =\left|  ^{\prime}off^{\prime}\right\rangle $
and $\left|  1\right\rangle =\left|  ^{\prime}on^{\prime}\right\rangle $,
where $\alpha_{1}$and $\alpha_{2}$ are the complex amplitudes of the clock
pulse basis states, and $\left|  \alpha_{1}\right|  ^{2}+\left|  \alpha
_{2}\right|  ^{2}=1$. The clock state \emph{induced} in the detector is:%

\begin{equation}
\left|  q\right\rangle \equiv\alpha_{1}\left|  0\right\rangle +\alpha
_{2}\left|  1\right\rangle
\end{equation}

The \emph{detected signal} is \emph{registered} as the \emph{direct product
}of the\emph{\ }$\emph{n}$\emph{-body energy eigenstates of the detector,
}$\left|  \Phi_{\gamma}\right\rangle $, with the\emph{\ induced excited state
}or\emph{\ phonon-like CE state, }$\left|  CE_{\gamma}\right\rangle $, of
the\emph{\ entire }$\emph{n}$\emph{-body system.} Examples of these kinds of
systems are the 'giant' multi-pole resonances of nuclei \cite{heyde} and
phonon behaviors (e.g. '\emph{Brillouin scattering'}) in crystals \cite{zim},
\cite{ssp}.

The configuration information of the mass-energy distribution is \emph{encoded
}in the state of the system which can be '\emph{measured}' to give a
conventional 'event time' label. The \textbf{excited }$\mathbf{n}%
$\textbf{-body state} of the component is given by:%

\begin{equation}
\left|  \Psi_{j}\right\rangle \equiv\left|  \Phi_{n-body}\right\rangle
{\LARGE \otimes}\left|  CE_{n-body}\right\rangle
\end{equation}

The \textbf{entangled triplet state} of this Feynman Clock is a composite
system of the three states above. The pulse counter 'labeled' and standard
clock pulse calibrated excited state is given by:%

\begin{align}
\left|  T_{j}\right\rangle  &  \equiv\left|  t_{j},q,\Psi_{j}\right\rangle
=\left|  t_{j}\right\rangle {\LARGE \otimes}\left|  q\right\rangle
{\LARGE \otimes}\left|  \Psi_{\gamma}\right\rangle \\
&  =\left|  t_{j}\right\rangle {\LARGE \otimes}\left(  \alpha_{1}\left|
0\right\rangle +\alpha_{2}\left|  1\right\rangle \right)  {\LARGE \otimes
}\left(  \left|  \Phi_{n-body}\right\rangle {\LARGE \otimes}\left|
CE_{n-body}\right\rangle \right)
\end{align}

which is an '\textbf{entangled }\emph{triplet state}' and cannot be reduced to
a simple linear sum of discrete states. It represents the entangled state of
the whole system which is the result of the system acting as a Feynman Gate
with 3 input signals and one output signal.

The disentanglement of these states occurs by a classical \emph{intervention}
\cite{Peres2}, \cite{Peres3}\ or \emph{measurement} of the entangled state
resulting in the extraction of the\ \emph{event time label} as a real number.
In the case of signal mapping, the processing of the triplet state signal
occurs in a 'gate' or FC that disentangles the event time from the other
information in the triplet state. The \textbf{disentanglement operator}, $D$,
acts on the triplet state via a \emph{classical intervention} causing the
decoherence of the coherent entanglement of the triplet state:%

\begin{equation}
D\left|  T_{j}\right\rangle =t_{e}\left|  t_{j},q,\Psi_{j}\right\rangle
\end{equation}

where $t_{e}$ is the \textbf{event time }corresponding to the \emph{classical
time} label for the moment of signal detection in the FD mode of the target FC
system. This is not the same as the \emph{lifetime} of an unstable state but a
'\emph{time label}'. The triplet state may someday be experimentally verified
in the actions on neural signals in microtubule causal networks in which
molecular conformation states represent the qubit states of 0 or 1, the
collective phonon resonance state of the microtubule represents the $n$-body
and its CE, and the counter label state is the number of neurotransmitters in
a chemical accumulator vesicle in the pre-synaptic membrane.

Recent work on the synchronization of quantum clocks provides a model for CEs
as entangled states in widely separated systems through a ''\emph{quantum
clock synchronization scheme}'' (QCS) \cite{qcent}. This model can be expanded
for Feynman Clock Synchronization (FCS) over 'classical' distances where the
FCs are virtual clocks (entangled 'time' independent \emph{signals}) until
'measured' or decohered from an atemporal global CE state into 'actual' FC
states of the nodes in a causal network. These synchronized nodes create a CEN
without the exchange of 'timing information'. Evidence of CEs over great
distances is found in photon entanglement experiments.

Experimental observation of two 'energy-time' entangled photons separated by
more than 10 Kilometers \cite{genFC} provides an example of the \emph{decay of
a collective excitation} of a vary large spatially extensive quantum system
\emph{if} we look at the entire experimental setup as a 'SEN' system from the
'Geneva FC' to the Bellevue/Bernex 'CEN'. The 'Geneva FC' produces two
'coherent' photon signals that traverse large distances on separate fiber
optic paths (8.1 and 9.3 km). The 'transit lifetimes' of the signals are
functions of the velocity of the signals in the medium and their distances to
the FDs in the Bellevue/Bernex CEN. Signal mapping of the FD/FC detection
events in the CEN via a 'clocked' memory system linking the two 'node' leads
to causal ordering. The entangled photons remained 'correlated' even though
separated by 10.9 kilometers, upon their detection 'decohere' with the
production of 'classical' information (i.e. the emission of 'signals' or the
creation of 'states' in memories) upon measurement.

The CEs of systems may act as measurements on the internal states by the
surface environment. This surface represents a plateau of complexity for these
systems. These plateaus have collective behaviors including irreversible
transitions to new configurations of matter and energy in expanding space. One
can artificially ascribe scaled arrows of time for these plateaus. These
system dependent arrows are derived from the quantum arrow of time. They
'\emph{correspond}' to the quantum arrow through the collective excitations
and behaviors of the networks of clocks and signals throughout the
hierarchical clusters of information processing subsystems.

\section{Collective Excitation Networks (CENs)}

Collective behaviors of systems composed of discrete but connected components
need to be characterized in order to understand how 'arrows of time' emerge at
POCs in complex systems. The concept of 'collective excitations' in the
many-body problem \cite{Mattuck} and in phonon behavior in solids \cite{zim},
\cite{ssp} provides the basis for modeling reconfigurations in POCs. When a
set of subsystems (local networks) in a complex system are 'wired' together in
a network, they can support coherent superposition of states capable of new
collective system behaviors. These collective states have finite lifetimes and
decay with the production of 'signals' (e.g. phonons, solitons, plasmons,
'sound waves', etc.).

The first level of complexity emerges when sets of \emph{coupled }Feynman
clocks act collectively as a single system with new system energy eigenstates
(e.g. molecular spectra) whose unstable excitation modes decay with finite
lifetimes. This system is a\textbf{\ Collective Excitation Network} or
\textbf{CEN}. These CENs can support new \emph{collective excitation states
and signals.} They can also act as 'gates', memories, or registers creating
and processing signals (information) when embedded in larger networks. This
process of 'nesting' of subsystems with collective excitation states provides
a means for deriving various hierarchical 'arrows of time' connected with
plateaus of complexity. Individual Feynman clocks and CEN units can interact
to form higher level CEN 'circuits'. These CEN circuits can become 'gates'
with \emph{multiple signal inputs and outputs}. These 'integrated' CEN
circuits now generate new POC states.

The 'lifetime' of the 'clock' mode of a general CEN is given by:%

\begin{equation}
\mathbf{\tau}_{CEN}=\frac{\hbar}{\mathbf{\Gamma}_{CEN}}=\frac{\hbar}{\left|
\langle\Psi_{CEN_{0}}\mid H_{CEN}\mid\Psi_{CEN}^{\ast}\rangle\right|  ^{2}}%
\end{equation}

where the excited 'clock' state of the CEN decays via the reconfiguration
transformation function, $H_{CEN},$ with the creation of a signal, $S_{out}$.
This is the 'lifetime' of a phonon resonance over a crystal array of atoms for instance.

The initial state of the CEN in the above equation is created by the detection
of an incoming signal, $S_{in}$ , by the CEN composed of a set of $j$-coupled
FCs. This 'system' configuration state, $\mid\Psi_{CEN}^{\ast}\rangle$, is the
\emph{direct product} of the states of each of the components:%

\begin{equation}
\mid\Psi_{CEN}^{\ast}\rangle=\left[  \bigotimes\limits_{i=1}^{j}\mid
\Psi_{FC_{i}}^{\ast}\rangle\right]  \bigotimes\mid\Psi_{S_{in}}\rangle
\end{equation}

The state of the CEN after decoherence ('decay' or 'decoupling') of the CE
over the set of FCs results in the emission of a signal, $S_{out}$ . The
'reconfigured' state of the system is:
\begin{equation}
\mid\Psi_{CEN_{0}}\rangle=\left[  \bigotimes\limits_{i=1}^{j}\mid\Psi_{FC_{i}%
}\rangle\right]  \bigotimes\mid\Psi_{S_{out}}\rangle
\end{equation}

The decohered FCs may still be bound in a lattice or other $n$-body
configuration ready to detect the next phonon-like signal.

\section{Sequential Excitation Networks (SENs)}

A SEN is a composite network of FCs and CENs coupled in such a way that
information and signals moves from node to node sequentially. The SEN has a
net 'lifetime' representing the sum of all the of the FC, CEN and signal
transit 'lifetimes from the initial signal input to a final signal output. The
SEN 'lifetime' for this process is given by:%

\begin{equation}
\mathbf{\tau}_{nsum}\equiv\sum\limits_{k}(\tau_{FC_{k}}+\tau_{S_{k}})
\end{equation}
where, $\tau_{FC_{k}}$, is the 'lifetime' of the $k$-th FC (or CEN) in the
sequence and $\tau_{S_{k}}$ is the signal lifetime between the $k$-th and
($k+1$) nodes.

\textbf{Feedback, feedforward }and\textbf{\ cyclical flow of signals
(information)} is also possible in the SEN. This provides a mechanism for the
resetting of unstable configurations necessary for quantum computational
algorithms. It also provides for adaptive behavior in relatively closed
systems like cells. These 'control' mechanisms can be realized by defining
signal trajectories or 'circuits' connecting various nodes into hybrid linear
and cyclical causal networks. All of the combinatorial possibilities for
'connecting' systems and subsystems together by signal loops provide a means
for modelling complex self-adjusting or adaptive behaviors. The
transformations of the local states or network configurations in the component
FD/FC, CEN, and SEN nodes produce different computational 'lifetimes' for the
information 'currents' propagating through them.

\section{Plateaus of Complexity (POCs)}

As we have seen above, collective excitations are the markers for new levels
of complexity in hierarchically connected systems. Solitons represent
'classical' wave packet signals in macroscopic scale systems. Their origins
are found in the \textbf{Plateaus of Complexity or POCs} of the subsystems
from which they are composed. Since CEs are the result of the superposition of
\emph{quantum states resulting in another quantum state,} classical states
emerge as the result of the interaction of this system with an
\emph{environment}. Plateaus of complexity are the interface between the
quantum properties of the system and its environment. This is how quantum
systems in CENs and SENs can create 'classical' signals and behaviors as a
result of the environmental measurement by an observing system in which it is
embedded. The environmental component makes the quantum system 'open' to
classical signal production. If the environment is the boundary condition on
the quantum system it may be 'closed', but still act like an open system which
can decohere (e.g. decay of FC mode of the initial state of the universe in
Big Bang scenarios).

POCs are configurations of complex systems from which \textbf{simplicity}
emerges. Simplicity of behaviors means that collective excitations provide a
way for classical physics to describe global changes without the need for a
complete description of the many individual systems contributing to the
overall 'simple' state. This is already evident in the success of classical or
Newtonian mechanics etc. POCs give rise to the behaviors and signals
associated with classical arrows of time and represent the basis for a
paradigm leading to a '\textbf{simplicity theory}' as a \emph{model} for the
\emph{emergence of hierarchical intermittent sets of simple POC states
punctuating the deterministic chaos intrinsic to} '\textbf{complexity
theory}'. Simplicity theory would then describe phenomena emerging as simple
large scale behaviors within complex systems.

\section{State Mapping and Processing into 'Time'}

\emph{''The Map is not the Territory''.-Alfred Korzybski }\cite{wopv3}

Signal mapping is the process by which signals carrying state information are
detected and their 'information content' (induced state in detector) put into
ordered sets with respect to a standard or internal clock. This involves
creating states in a 'memory' so that their causal relation to other events
can be 'read' and interpreted. 'Time' emerges as the functional value of the
energy eigenstates in the detectors as information 'bits' assigned to a
detected signal from an 'event' (FC created signal) in $3$-space (possibly an
$n$-space at the Planck scale for $n$ higher dimensional quantum modes of
'strings' etc.). The magnitude of the states (in 'bits') are determined by the
conversion of state information by a detector and kept in a memory register as
a mirror state of the original source state created by the decay of the signal
generating FC. The state in the memory can be 'scanned' (measured) by a shift
or parallel data register through the action of an internal or standard clock.
This is similar to data ordering in classical computational hardware.

The process of signal or state mapping resulting in a 'time' number label for
a signal induced state in memory is the result of \emph{processing} a
'triplet' state. The creation of this entangled state results in a new state
that encodes a label referenced to a standard clock register for the 'time of
detection' for the creation of an FD state:%

\begin{equation}
\mid T_{j}\rangle=\mid t_{j}\rangle\bigotimes\mid1\rangle\bigotimes\left|
\Psi_{j}\right\rangle =\mid t_{j},1,\Psi_{j}\rangle
\end{equation}

A 'disentanglement' measurement, $D$, on this state gives the number
representing the 'time' label for the detection of the signal event with
respect to a standard clock (e.g. cyclical FC powered by regular signals from
the 'environment'). The measurement gives the 'time label' eigenvalue $j $ in
the equation below:%

\begin{equation}
D\mid T_{j}\rangle=D\mid t_{j},1,\Psi_{j}\rangle=t_{e}\mid t_{j},1,\Psi
_{j}\rangle
\end{equation}

This value is the 'classical' time for the event where $t_{e}\in\Re$ (set of
real numbers). The disentanglement of this triplet state requires a quantum
computer gate that 'selects' the 'label' state and can transfer this extracted
information to a symbol or word in a language to be communicated to other
systems. The exact physical situations that can produce such entanglement and
disentanglement will be explored in future work.

The key point here is that all of the systems (FC, Signal, FD, Memory and
cyclical data sequencing clock) may be 'quantum' systems with microscopic or
classical sizes. In this way, the relative order and magnitude of the
conventional 'time' interval \emph{between events} is the result of the
processing of state information in the 'gates' of a \emph{quantum computer}.

\section{Examples}

\subsection{The 'Big Bang' as a Feynman Clock}

Can the Big Bang Singularity be modeled as a FC? The 'diameter' of the FC
universe at that point was of the order of the \emph{Planck Length} or about
$1.61605\times10^{-35}%
\operatorname{m}%
$ indicates that the methods of quantum mechanics are appropriate for the
description of the Big Bang initial state \cite{DeWitt}, \cite{kolb}. The
Wheeler-DeWitt Equation is a starting point for 'quantum cosmology'. The
inflationary Big Bang scenarios are designed to accommodate an extremely high
energy density singularity \cite{asp}, \cite{hawking}, \cite{kolb},
\cite{linde}, \cite{guth} as the source for all structure observed today.

The initial 'excited' unstable state of the universe is a singular unstable
high energy-density Feynman Clock with '\textbf{decay products}' \emph{such
as} the '\emph{vacuum}' (space), various forms of matter, the fundamental
interactions between particles, and evolving complex systems in causal
networks. The emergence of the 'vacuum' at the end of inflation creates an
'environment' which decoheres the previously coherent pre-inflation
superposition of configuration states from the continuum of all possible
('future') coupled decay states for the initial singularity \cite{zizzi3}.

The initial unstable state is the direct product of all the coherent modes of
a 'pseudo-stable state' coupled to a global collective excitation in the form
of a phonon-like perturbation in the mass-energy density function of the
initial singularity. This phonon-like perturbation is frozen out into
spatially distinct mass-energy density enhancements when the vacuum 'signal'
decouples from matter at the end of the inflationary epoch resulting in the
first Feynman Clock '\emph{tick}' of the Universe. These density enhancements
evolve into complex systems through '\textbf{quantum'} \textbf{source and
'classical' sink} \emph{non-equilibrium competition} between the strong,
electromagnetic, and weak interactions of matter in the continuously
\emph{stretching} vacuum \textbf{versus} the local gravitational clustering of
mass in the form of galaxies, stars, and planets. The gravitational clustering
of matter is essential for the emergence and evolution of complex systems of
matter (e.g. the ecosphere and the life within it) driven by energy sources
such as stars.

We propose that an initial 'triplet' state of the Universe is modeled with an
entangled state built from the \textbf{coherent set of FC configuration
states} $\mid\Psi_{i}\rangle$, the \textbf{global collective excitation state}
$\left|  \Psi_{CE}\right\rangle $ over this system \emph{and} the
\textbf{Planck Scale FC counter state} $\left|  t_{U}\right\rangle $. The
resultant initial entangled triplet FC state of the non-tunneling decay mode
of the Universe is:%

\begin{equation}
\mid\Psi_{U_{i}}\rangle=\left(  \bigotimes\limits_{i=1}^{\infty}\mid\Psi
_{i}\rangle\right)  \bigotimes\left|  \Psi_{CE}\right\rangle \bigotimes\left|
t_{0}\right\rangle
\end{equation}

decohering via 'self-measurement' from the CE 'Environment'. This forces a
'phase transition' producing an inflationary Big Bang FC evolutionary causal
network in the vacuum with a \textbf{first cause} decay mode 'decoherence
lifetime' of:
\begin{equation}
\mathbf{\tau}_{BigBang}=\frac{\hbar}{\mathbf{\Gamma}_{U_{FC}}}=\frac{\hbar
}{\left|  \langle\Psi_{U_{Inflation}}\mid D_{U_{PlanckTime}}\mid\Psi_{U_{i}%
}\rangle\right|  ^{2}}=\tau_{PlanckTime}=5.39056\times10^{-44}%
\operatorname{s}%
\end{equation}

where:%

\begin{equation}
\left|  \Psi_{U_{Inflation}}\right\rangle =\left(  \bigotimes\limits_{k=1}%
^{\infty}\mid\Psi_{k}\rangle\right)  \bigotimes\left|  \Psi_{Vac}\right\rangle
\bigotimes\left|  t_{P}\right\rangle
\end{equation}

and $D_{U_{PlanckTime}}$ is the \textbf{time-independent } FC-Universe
reconfiguration operator. The 'energy of reconfiguration' term, $\langle
\Psi_{U_{Inflation}}\mid D_{U_{PlanckTime}}\mid\Psi_{U_{i}}\rangle$, is
encoded in the '\textbf{expanding universe signal}' seen as the creation of a
\emph{global expanding vacuum environment} with \emph{gravitationally created
FC systems} formed by the clustering of matter and energy into \emph{local}
information 'sources'. As this mass-energy-information density 'locally'
increases in the form of 'sources' such as galaxies, stars, planets and
humans, we see that the density of 'signals' and therefore available
information (states) forming causal links between these spatially distinct
systems decreases as particle, atomic, and molecular \emph{FC density} of
space declines due to expansion of the universe. The increase in local POCs is
in stark contrast with the increasing unavailability of energy sources in an
expanding universe.

The fundamental interactions between particles emerge from the destruction of
the coherence of configuration states as the 'interference terms' between
topological inhomogeneities in the energy density function. The evolution of
the FC-universe into a hierarchy of complex systems of causal networks forming
a 'quantum computer' is a topic for further speculation \cite{lloyd}. The
quantum computer analogy may be explored by taking the position that the
initial early universe was a FC or CEN that decohered to a configuration of
matter and energy that then 'decayed' in an inflationary SEN of branching,
subdividing, and hierarchically connected FC, CEN and SEN 'gates'.

The continuous evolution and \emph{branching} of causal networks of matter and
signals makes it difficult to treat the universe as a 'single' quantum
computer system representing all of the emergent structures in the universe
throughout its reconfiguration history. The branchings are in one system and
do not need the many \ distinct universes to 'branch into'. This does not mean
that the universe itself is not a 'quantum computer', but that it might be a
FC-quantum computer which can accommodate the complex hierarchical causal
networks and the signal mediated information flow in them evolving into
differentiated 'classical' structures in which the quantum component systems
are subsumed.

\subsection{'Unification' of the Fundamental Interactions}

Is it possible to \emph{unify} the strong, electromagnetic, weak and
gravitational forces using 'time' as a common term? Any one or a combination
of the strong, electromagnetic, weak and gravitational fundamental
interactions can drive reconfiguration processes in FCs, CENs and SENs. In
this sense all of these interactions have 'lifetimes' and therefore
information generating capabilities in common and are therefore 'unified' in
an \emph{information space} \cite{infospace}. For a FC reconfigured by the
strong interaction we have a decay or decoherence lifetime $\mathbf{\tau}_{U}
$:%

\begin{equation}
\mathbf{\tau}_{U}=\alpha\mathbf{\tau}_{strong}=\frac{\hbar}{\mathbf{\Gamma
}_{strong}}%
\end{equation}

For a FC system driven by the weak interaction (or 'electroweak') we have:%

\begin{equation}
\mathbf{\tau}_{U}=\beta\mathbf{\tau}_{weak}=\frac{\hbar}{\mathbf{\Gamma
}_{weak}}%
\end{equation}

For a FC system driven by the electromagnetic interaction we have:%

\begin{equation}
\mathbf{\tau}_{U}=\delta\mathbf{\tau}_{em}=\frac{\hbar}{\mathbf{\Gamma}_{em}}%
\end{equation}

and for a gravitational FC system we have:%

\begin{equation}
\mathbf{\tau}_{U}=\epsilon\mathbf{\tau}_{grav}=\frac{\hbar}{\mathbf{\Gamma
}_{grav}}%
\end{equation}

where the lifetimes are related by real scalar constants $\alpha$, $\beta$,
$\delta$, and $\epsilon$. The unified 'lifetime', $\mathbf{\tau}_{U}$ is then:%

\begin{equation}
\mathbf{\tau}_{U}=\alpha\mathbf{\tau}_{strong}=\beta\mathbf{\tau}%
_{weak}=\delta\mathbf{\tau}_{em}=\epsilon\mathbf{\tau}_{grav}%
\end{equation}

These four prototypical systems are reconfigured by different forces but their
signals provide a rather obvious and perhaps trivial way of establishing an ad
hoc unification of the fundamental interactions of matter. The key to this
type of unification is recognizing the dimensional equivalence of the
'lifetimes' and therefore the \textbf{reconfiguration information}
\emph{common} to all unstable systems. Note that these are the 'QAT lifetimes'
of states in FC systems. The signals carrying information from two different
types of the FCs above to a detection systems or observer are converted into
CAT differences between the detector events.

\subsection{Time Travel?}

The interface between the quantum basis of the classical world along with the
irreversible nature of the QAT and the reversible CAT brings up the question
of whether '\textbf{Time Reversal' }or\textbf{\ 'Time Travel'} is
possible\textbf{. }The popular conception of time travel implies that
'\textbf{past'} and '\textbf{future'} configuration states of the universe
coexist and are somehow accessible from the '\textbf{present'. }If the
\textbf{essential structural information} about the 'present' state of a FC is
lost in the form of \textbf{signals} to the \textbf{environment} during a
reconfiguration then the \textbf{resetting }of the decayed state by incoming
signals \textbf{appears} to constitute a 'time reversal' of that FC. The
\textbf{'reset' signals} with the essential information necessary to
reconfigure the FC for this \textbf{apparent reversibility }comes from the
\emph{classical} environment.

This may create a \emph{false sense} that the local properties of a system
have been '\textbf{returned}' to a previous unstable state that somehow also
coexists with the 'present state' of the observer or 'Time Traveler'. The
environment (e.g. universe) of this system however \emph{has not} undergone a
return to an identical previous unstable state due to its global evolution.
The environment state is not consistent with the 'previous' local FC state due
to the increase in\emph{\ information entropy of the combined system of local
FC and the rest of the universe}. This is a restatement of the second law of
thermodynamics and the origins of entropy at the microscopic level from
quantum sources to the statistical ensembles of classical mechanics.

The internal QAT for this FC is always \textbf{irreversible}. The unstable
state that has been created by incoming signals decays \textbf{irreversibly}
even though the 'reset signal' is a pointer from the environment to the
quantum system indicated by a reversible CAT coupled to the trajectory or
direction of the signal in space. It must be noted that the information used
to create this unstable state has been lost by another system. The combination
of the \textbf{expansion of space} \emph{and} the \textbf{gravitational
condensation of matter} into galaxies, stars and planets maps the overall
decrease in availability of the information necessary to \emph{'re'-create}
the unstable configuration state required for \textbf{backward time travel}.
'\textbf{Forward' time travel} only requires that one's \emph{clocks} be
\textbf{slowed} \textbf{down }(i.e. larger 'energies of reconfiguration'
giving longer 'lifetimes' for transitions between configurations) by either
relativistic velocities, cryogenic slowing of metabolism, or modification of
DNA to slow or reverse ageing processes. The '\textbf{future}' will be the
\textbf{result} of the evolutionary processes and information flow in
hierarchical causal networks. It is meaningless to talk about the outcomes of
these processes as existing simultaneously in parallel or among branching
multiple universes since they have yet to be created. Time travel' and 'time
reversal' are not possible in a global sense since this requires modification
of the subject system or FC to a previous or future state by reversed or
forward information flow in signals from the environment in which it is
embedded. The paradox is evident in the concurrent necessity for the
environment to also be reconfigured to a 'previous' state identical with the
one from which the original FC system decayed!

\subsection{Time 'Dilation' and 'Contraction'}

\emph{Time intervals between events} (CATs) and the \emph{'lifetimes' of
unstable systems} (QATs) are subject to '\emph{dilation}' and
'\emph{contraction}'. These correspond to two cases: 1) relativistic motion
and energy effects, and 2) \textbf{Quantum Zeno Effects} \cite{asher} for
'classical' (including 'relativistic') and 'quantum' systems respectively.

For the 'classical' case of dilation or contraction of \emph{time intervals}
where relativistic corrections for source-detector motion are required, the
signals from two space-like separated 'moving' FCs (or a \emph{second signal}
from the same FC) may appear to be increased or 'dilated' for objects moving
away from the observer. They may appear to be decreased or 'contracted' for
motion towards the observer. These signals may originate as \emph{quantum
processes} in the FCs but they are\emph{\ processed} ('signal mapping') as two
separate detector triplet state events in the observer system. The
\emph{triplet states }created in the detectors are '\emph{time labeled}' and
the computation of the \emph{differences} between two or more of these 'event
times' using the standard or internal clock of the observing system is subject
to appropriate relativistic corrections depending on the geometry and motion
of the sources with respect to the observer. The relativistic corrections are
\emph{computed} after signal processing and 'time difference' maps have been
created in a 'memory'.. These 'computed' relativistic '\emph{time
dilation/contraction'} effects between detection events represent an
'\emph{interpretation' process} which is a \emph{map in information space
}that gives the causal relationships of the sources and the detectors and
based on other information about the systems whether the signal time
separations were due to their geometric relationships or other local
processes. This is necessary to 'understand' whether relativistic phenomena
have been observed based on the detected signals.

The case of dilation or contraction of 'lifetimes' of unstable states in
quantum systems refers to the modification of the intrinsic decay process by
interventions inducing more or less stability in the state. More stability
means that the system will stay together longer before decay thus dilating the
'intrinsic lifetime'. More instability contracts the intrinsic lifetime and
shortens the life of the unstable state. It should be noted that both dilation
and contraction effects of the quantum lifetime are really the lifetimes of
\emph{new hybrid entangled states} \ in a \emph{composite system} of
environment (e.g. incoming signals) and the FD mode of a FC. This is the
quantum Zeno effect which can act to slow or speed up a decay process
depending on the nature of the interaction of the environment with the quantum
system. The interventions of drugs upon the brain for instance can induce time
'dilation' and 'contraction' effects on the \emph{consciousness CE state} by
'\emph{spreading out}' or '\emph{squeezing' }the CE state respectively (see
below). Dilation of this state means that more signals than normal per CE
state lifetime are processed thus creating the sensation that the world is
moving faster. Contraction of the CE state means that less signals than normal
are processed per CE lifetime and therefore the rapid sequence of CE states
creates the appearance that the world is moving slower. this is the case for
the 'adrenaline rush' experience of people in crisis situations who
experiences time slowing down. The use of chemicals that trigger Quantum Zeno
effects in neurons may be present the possibility for
experimental\ verification of CEN quantum states in neural networks such as
the brain.

\subsection{FTL Signals and Superluminal Information Flow}

Faster than light (FTL) signals have been 'observed' \cite{ftl} in systems
that appear to produce causal effects before their trigger signals are
completely detected. While the results are preliminary, it may be that the
speed of information \cite{sqi} is 'superluminal' for special cases of
entangled states whose \emph{disentanglement} is confused with speeds of
\emph{electromagnetic} signals exceeding that of light in a vacuum. The
decoherence of a CE over two or more time-like separated FC entangled states
in a CEN creates classical states whose causal network interactions appear to
be superluminal but are really the result of the space between nodes acting as
a \textbf{pseudo-FC} whose decay lifetime is less than the lifetime of a
signal traversing the distance between FCs at the speed of light. This
pseudo-FC can also be thought of as the \emph{process} of '\textbf{signal
tunneling}' through 2 '\emph{barriers'.} The barriers are the FC and FD node
endpoints forming the two energy levels ('1' or 'excited' and '0' or 'ground'
respectively) of a two state pseudo-FC. The signal 'travels' with infinite
speed \cite{tunnel} through the internodal vacuum providing a possible
mechanism for the non-local synchronization of states necessary for the
emergence of spatially extended CEs.

\subsection{The Emergence of Consciousness}

How does \emph{consciousness} emerge in the complex neural networks of the
brain? There is some question about whether the Brain needs to be modeled as a
quantum computer or system \cite{qbrain}, \cite{mershin}. It appears that the
brain has both quantum and classical information generating and processing
properties depending on the size of the subsystem under examination.
\emph{Coherence} of the many individual states of neurons seems to be
necessary for the creation of a single large scale 'thought'. The coherence
mechanism may be classical such as the release of energy stored in a
\textbf{many-body} neural network creating phonon-like CE 'brain waves'
\cite{ye}. This however ignores the possible complex interactions of the
internal structures of neurons at the quantum level \cite{alex},
\cite{ceneuron}, \cite{chemwaves}. It has been proposed that \textbf{quantum
gravity }is involved with consciousness and quantum computational processing
of 'holographic' patterns of information at the Planck scale \cite{hht},
\cite{zizzi}, \cite{zizzi2}. It is likely that 'consciousness' is the result
of both quantum and classical information processing and storage.

At the neuron level, microtubule structures in the cytoskeleton may act as
\textbf{qubit} \cite{eiqc}, \cite{phys229}\ processing gates in a molecular
quantum computer \cite{hht}. The resulting states in these neurons may be
synchronized with other neurons in a neural network by entangled states
mediated by internal photon and external phonon-like collective excitations.
They may also be synchronized by classically mediated neurotransmitter release
and uptake at pre- and post synaptic membrane sites via diffusion and
electro-chemistry. Macroscopic electromagnetic properties seen in brainwaves
and NMR images of active regions of the brain may not need a quantum
explanation but yet may still be the result of classical POC collective
excitation states resulting from the synchronization of the quantum chemical
causal networks with FC, CEN and SEN information processing gates. The quantum
and classical properties of the brain giving rise to 'consciousness' may meet
at the mesoscopic realm of the synaptic gap. The neuron may be a quantum
computer connected to many other quantum computers forming a classical
computer neural network.

There are many open questions in this difficult and complex problem. What we
propose here is that QATs in neurons may give rise to CATs in neural networks
through hierarchical POCs. If time is a form of information then the
information processing capabilities of neurons (or 'gates') in networks may be
required for the ordering and time labeling of events in the world around us.
The exploration of the fundamental nature of time leads naturally to the
problem of consciousness.

If neurons are local quantum computers in networks then the irreversible
processes in them should give rise to QATs pointing along the paths of
'information flow'. If microtubules in neurons can be \emph{synchronized} into
'single' collective excitation states either by entanglement of signals
coupled by \textbf{faster than light 'information' communication }\cite{ftl}
or by a \textbf{vacuum induced entanglement} \cite{benni} then there is the
possibility for a quantum description of 'thoughts' as classical
electromagnetic phenomena emerging from spatially extended entangled quantum
systems such as those in neural networks of the brain.

A recent analysis of EPR experiments in Geneva indicates a lower bound of
$\mathbf{1.5\times10}^{4}\mathbf{c}$ for '\textbf{communication}' between two
\emph{entangled} photons with space-like separations of $10.6$ $%
\operatorname{km}%
$ \cite{sqi}. This may pave the way for understanding how photons and phonon
like resonances in individual and network microtubules might support a
'quantum state' of consciousness. Less spectacularly, resonance effects of
large scale classical electromagnetic 'brainwaves' may coordinate the quantum
states. The quantum or classical CE states of neural networks emerge and decay
with characteristic 'lifetimes'. The superposition of these 'local' CEs
generated by overlapping CENs can synchronize even larger sets of neurons into
a large scale CE. This may give rise to the distributed collective state over
a large set of neurons supporting a singular but continuously transforming
'consciousness' state.

The 'lifetime' of any given 'thought' as a net CE state configuration
resulting from the transformation $B_{1\rightarrow2}$ of a decaying CE;
$\Psi_{CE_{1}}$ to another overlapping emergent CE; $\Psi_{CE_{2}}$ forms a
SEN of \textbf{POC states of consciousness} is:%

\begin{equation}
\mathbf{\tau}_{SEN_{1,2}}=\frac{\hbar}{\mathbf{\Gamma}_{SEN_{1,2}}}%
\end{equation}

The brain is 'quantum' only in the sense that complex states (e.g.
consciousness) can be viewed as collective excitation state of casual networks
built from neuron based Feynman Clocks, CENs, and SENs. The building of these
complex interacting neural networks give rise to 'classical' global collective
excitations such as brain waves. The collective excitation states in the brain
may be created by the adaptive rewiring of some neurons among relatively fixed
function neurons allowing a continuity of historical memory information while
processing new information generated by the\ sequential emergence and decay of
novel CEN states. The ultimate 'collective excitation' may be consciousness.
As far as we know it is the Brain that maps numbers onto events and constructs
the dimension and direction of time from a changing world that cares not for
the numbers we use to define our lives.

\section{Summary}

It has been postulated that \textbf{time is a form of information}.
Information ultimately takes the form of labels, words, and language in
complex systems such as the brain. The source of all information is the
initial unstable collective excitation state of the Feynman Clock Universe.
The purpose of the theory presented above is more 'explanatory' than
'predictive'. The 'triplet' state necessary for the creation of 'time' labels
for the signal induced events in detectors or memory registers may eventually
be 'observed' in the microstructure of neurons. The understanding of the
fundamental nature of time as information rather than as it's
\emph{constructed} 'dimension' and 'direction' may lead to resolution of
'time' related causality paradoxes in quantum mechanics and relativity theory.

The information processing properties of complex configurations of matter and
energy in quantum computers may lead to an understanding of the emergence of
the brain and it's most powerful state- 'consciousness'. The flow of
information in the form of signals between emergent hierarchical patterns of
matter map the evolution of the universe. It is hoped that the novel ides,
models and tools presented in this paper will help contribute to the answers
to the 'age' old questions about the origins of the universe and how a 'brain'
could emerge from the Big Brain...\emph{after all}, \textbf{understanding the
Brain is the ultimate test for any theory of 'everything', 'anywhere' at 'anytime'!}

\section{Acknowledgments}

I am grateful to \textbf{Professors Vladimir A. Petrov, Anatoly A. Logunov,
Sergei Klishevich, and the Organizing Committee of the XXIII International
Workshop on the Fundamental Problems of High Energy Physics and Field Theory},
June 21-23, 2000, for their generous invitation to share these rather novel
and complex set of ideas with them at the \textbf{Institute for High Energy
Physics}, Protvino, Moscow Region, Russia. I would like to thank
\textbf{Professor} \textbf{Arkadi I. Lipkine} (MPTI, Moscow) for insights and
critiques of my work. I would also like to thank \textbf{Professors}
\textbf{Gert G. Harigel, Guang-Jiong Ni, L. Smekal,} and \textbf{A. Mitra
}for\textbf{\ }valuable discussions, cross-examinations, and 'reality'
checks\textbf{.} It has been a great privilege to meet all the other
participants of this stimulating and well organized Workshop.

A special thanks to my wife \textbf{Toni} for sharing this adventure with me.
I would like to thank \textbf{James W. Sterling} for designing and building
the \emph{'Feynman Clock Demonstration Units' }used in my presentation. Note:
some of the references below were obtained during the preparation of this
paper and are included for completeness.


\begin{thebibliography}{99}
\bibitem{hitchcock}''\textbf{Quantum Clocks and the Origin of Time in Complex
Systems}'' by \emph{Scott Hitchcock}, LANL e-print archives; gr-qc/9902046 v2,
20 Feb 1999, also NSCL Publication: MSUCL-1123, 1999. \textbf{Note: for all of
the following references to ''LANL e-print archives...'', they can be viewed
and downloaded at the Los Alamos National Laboratory web page at http://xxx.lanl.gov/.}

\bibitem{hitchcock2}''\textbf{Feynman Clocks, Causal Networks, and the Origin
of Hierarchical 'Arrows of Time' in Complex Systems from the Big Bang to the
Brain. Part I 'Conjectures'''}by \emph{Scott Hitchcock}, LANL e-print
archives; gr-qc/0005074, 16 May 2000, also NSCL Publication: MSUCL-1135, 2000.

\bibitem{Peres}''\textbf{Measurement of time by quantum clocks}'' by
\emph{Asher Peres}, Am. J. Phys. 48(7), July 1980.

\bibitem{asher}''\textbf{Quantum Theory: Concepts and Methods'' }by
\emph{Asher Peres}, Kluwer Academic Publishers, Dordrecht, the Netherlands, 1998.

\bibitem{eiqc}''\textbf{Explorations in Quantum Computing'' }by \emph{Colin P.
Williams }and \emph{Scott H. Clearwater}, Springer-Verlag, New York, 1998.

\bibitem{phys229}''\textbf{Lecture Notes for Physics 229: Quantum Information
and Computation''} by \emph{John Preskill}, California Instituter of
Technology, September 1998, available at
http://www.theory.caltech.edu/\symbol{126}preskill/ph229.

\bibitem{davies}''\textbf{About Time, Einstein's Unfinished Revolution'' }by
\emph{Paul Davies}, Simon and Schuster, NY, 1995.

\bibitem{timeless}''\textbf{Timeless'' }by \emph{Julian Barbour}, New
Scientist, 16 October 1999, pp. 29-32.

\bibitem{marchal}''\textbf{Determinism and Arrow of Time: Henri Poincar\'{e}
Philosopher and Scientist''} by \emph{C. Marchal}, \textbf{Fundamental
Problems of High Energy Physics and Field Theory, Proceedings of the XXII
International Workshop on High Energy Physics and Field Theory}, Protvino,
Moscow Region, June 23-25, 1999, Editors\emph{\ I. V. Filimonova }and \emph{V.
A. Petrov}, State Research Center of Russia, Institute for High Energy
Physics, pages 108-120.

\bibitem{prig}''\textbf{Order Out of Chaos''} by \emph{Ilya Prigogine
}and\emph{\ Isabelle Stengers}, Bantam Books, New York, 1984.

\bibitem{zeh}''\textbf{The Physical Basis of The Direction of Time''} by
\emph{H. D. Zeh}, Springer-Verlag, 3rd Edition, Berlin, 1999.

\bibitem{cove}\textbf{''The Arrow of Time, A voyage through science to solve
time's greatest mystery''} by \emph{Peter Coveney }and \emph{Roger Highfield},
Fawcett Columbine, NY 1990.

\bibitem{schulman1}''\textbf{Time's arrows and quantum measurement''} by
\emph{L. S. Schulman}, Cambridge University Press, United Kingdom, 1997.

\bibitem{rirnr}\textbf{''Reversibility, irreversibility: restorability,
non-restorability''} by \emph{B. Bernstein }and\emph{\ T. Erber}, J. Phys. A:
Math. Gen. 32 (1999) 7581-7602.

\bibitem{dacwqt}''\textbf{Decoherence and the Appearance of a Classical World
in Quantum Theory''} by \emph{D. Giulini, E. Joos, C. Kiefer, J. Kupsch, I.-O.
Stamatescu, }and\emph{\ H. D. Zeh}, Springer-Verlag, Berlin, 1996.\textbf{''}

\bibitem{zurek}\textbf{''Decoherence and the Transition from Quantum to
Classical''} by \emph{Wojciech H. Zurek}, Physics Today, October 1991, pp. 36-44.

\bibitem{Diu}''\textbf{Quantum Mechanics}'' Volume Two, by \emph{Claude
Cohen-Tannoudji, Bernard Diu, and Franck Lalo\"{e}}, John Wiley and Sons,
France, 1977. See pages 1344-1356 for the basic model used to develop the
Feynman clock in this paper. \textbf{The key to the theory outlined in this
paper is the irreversible coupling of an unstable discrete state of a system
to a 'continuum' of possible discrete reconfiguration states of the system
from which one is 'chosen' as a result of the decay or 'decoherence' process
('self-measurement')}. This fundamental mechanism is expanded for use in the
development of a systems approach involving collective excitations of coupled
sets of clocks.

\bibitem{naqw}''\textbf{Nonlocal Aspects of a Quantum Wave''} by \emph{Y.
Aharonov }and\emph{\ L Vaidman, }LANL e-print archives; arXiv:
quant-ph/9909072, 23 Sep 1999.

\bibitem{qdae}''\textbf{Quantum decoherence from Adiabatic Entanglement''} by
\emph{C. P. Sun, D. L. Zhou, S. X. Yu, }and\emph{\ X. F. Liu}, LANL e-print
archives; arXiv: quant-ph/0001068, 19 Jan 2000.

\bibitem{hologram}\textbf{''A quantum holographic principle from
decoherence''} by \emph{Sougato Bose and Anupam Mazumdar}, LANL e-print
archives; arXiv: gr-qc/9909008, 2 Sep 1999.

\bibitem{ftl}''\textbf{Gain-assisted superluminal light propagation''} by
\emph{L. J. Wang, A. Kuzmich }and\emph{\ A. Dogariu}, Nature, Volume 406, 20
July 2000, Pages 277-279.

\bibitem{sqi}''\textbf{The Speed of Quantum Information and the Preferred
Frame: Analysis of Experimental Data'' }by \emph{Valerio Scarani, Wolfgang
Tittel, Hugo Zbinden, }and\emph{\ Nicolas Gisin}, LANL e-print archives;
quant-ph/0007008, 4Jul 2000.

\bibitem{tunnel}''\textbf{Tunneling through two successive barriers and the
Hartman (Superluminal) effect''} by \emph{Vladislav S. Olkhovsky, Erasmo
Recami, }and\emph{\ Giovanni Salesi}, LANL e-print archives; arXiv:
quant-ph/0002022 v3, 18 Aug 2000.

\bibitem{lloyd}\textbf{''Universe as quantum computer''} by \emph{Seth Lloyd},
LANL e-print archives, quant-ph/9912088, 17 Dec 1999.

\bibitem{Veltman}''\textbf{Diagrammatica; The Path to ''Feynman Diagrams}'' by
\emph{Martinus Veltman}, Cambridge University Press, 1995.

\bibitem{Mattuck}''\textbf{A Guide to Feynman Diagrams in the Many-Body
Problem}'' by \emph{Richard D. Mattuck}, Dover Publications, Inc., New York, 1992.

\bibitem{zago}''\textbf{Quantum Theory of Many-Body Systems}'' by
\emph{Alexandre M. Zagoskin}, Springer-Verlag, New York, 1998.

\bibitem{heyde}\textbf{''Basic Ideas and Concepts in Nuclear Physics''} 2nd
Edition, by \emph{K. Heyde}, Institute of Physics Publishing, London, UK, 1999.

\bibitem{zim}''\textbf{Principles of the Theory of Solids''} by \emph{J. M.
Ziman}, Cambridge University Press, Great Britain, 1995.

\bibitem{ssp}''\textbf{Solid State Physics''} 2nd Edition, by \emph{J. S.
Blakemore}, W. B. Saunders Company, Philadelphia, Pennsylvania, 1974.

\bibitem{Peres2}''\textbf{Classical interventions in quantum systems. I. The
measuring process}'' by \emph{Asher Peres}, LANL e-print archives;
quant-ph/9906023, 7 June 1999.

\bibitem{Peres3}''\textbf{Classical interventions in quantum systems. II.
Relativistic invariance'' }by \emph{Asher Peres}, LANL e-print archives;
quant-ph/9906034, 10 June 1999.

\bibitem{qcent}\textbf{''Quantum Atomic Clock Synchronization Based on Shared
Prior Entanglement''} by \emph{Richard Jozsa, Daniel S. Abrams, Jonathan P.
Dowling, }and\emph{\ Colin P. Williams}, LANL e-print archives; arXiv:
quant-ph/0004105, 27 Apr 2000.

\bibitem{genFC}''\textbf{Long-distance Bell-type tests using energy-time
entangled photons}'' by \emph{W. Tittel, J. Brendel, N. Gisin, and H.
Zbinden}, Physical Review A, volume 59, Number 6, June 1999.

\bibitem{wopv3}''\textbf{The World of Physics Volume III''} by \emph{Jefferson
Hane Weaver}, Simon and Schuster, New York, 1987.

\bibitem{DeWitt}''\textbf{Quantum Theory of Gravity. I. The Canonical
Theory}'' by \emph{Bryce S. DeWitt}, Physical Review, Volume 160, Number 5, 25
August 1967, pp. 1113-1148.

\bibitem{asp}''\textbf{From Quantum Fluctuations to Cosmological Structures}''
\emph{Edited by David Valls-Gabaud, Martin A. Hendry, Paolo Molaro, and Khalil
Chamcham}, Astronomical Society of the Pacific, San Francisco, CA, 1997.

\bibitem{hawking}''\textbf{The Nature of Space and Time}'' by \emph{Stephen
Hawking} and \emph{Roger Penrose}, Princeton University Press, USA, 1996.

\bibitem{kolb}''\textbf{The Early Universe}'' by \emph{Edward W. Kolb} and
\emph{Michael S. Turner}, Addison-Wesley Publishing Company, USA 1994.

\bibitem{linde}''\textbf{Particle Physics and Inflationary Cosmology}'' by
\emph{Andrei Linde}, Harwood Academic Publishers, The Netherlands, 1996.

\bibitem{guth}''\textbf{The Inflationary Universe}'' by \emph{Alan H. Guth},
Addison-Wesley, Reading, Massachusetts, 1997.

\bibitem{infospace}''\textbf{Classical and quantum mechanics on information
spaces with applications to cognitive, psychological, social and anomalous
phenomena''} by \emph{Andrei Khrennikov}, LANL e-print archives;
arXiv:quant-ph/0003016, 4 Mar 2000.

\bibitem{zizzi3}''\textbf{Quantum Computation Toward Quantum Gravity''} by
\emph{Paola A. Zizzi}, LANL e-print archives; arXiv: gr-qc/0008049, 21 Aug 2000.

\bibitem{qbrain}''\textbf{The Importance of Quantum Decoherence in Brain
Processes''} by \emph{Max Tegmark}, LANL e-print archives, quant-ph/9907009
v2, 10 Nov 1999.

\bibitem{mershin}''\textbf{Quantum Brain?}'' by \emph{Andreas Mershin, Dimitri
V. Nanopoulos, }and\emph{\ Efthimios M. C. Skoulakis}, LANL e-print archives;
quant-ph/0007088, 24 Jul 2000.

\bibitem{ye}\textbf{''Coherence in a simple network: Implication for brain
function}'' by \emph{Zhen Ye}, LANL e-print archives; cond-mat/0007242, 14 Jul 2000.

\bibitem{alex}''\textbf{Hierarchic Model of Consciousness: From Molecular Bose
Condensation to Synaptic Reorganization'' }by \emph{Alex Kaivarainen}, LANL
e-print archives; arXiv: physics/0003045, 21 Mar 2000.

\bibitem{ceneuron}''\textbf{Modeling collective excitations in cortical
tissue'', }by \emph{Werner M. Kistler, Richard Seitz, }and\emph{\ J. Leo van
Hemmen}, Physica D 114 (1998) 273-295.

\bibitem{chemwaves}''\textbf{From Neurons to Brain: Adaptive Self-Wiring of
Neurons''} by \emph{Ronen Segev }and\emph{\ Eshel Ben-Jacob}, LANL e-print
archives; arXiv: cond-mat/9806113, 9 Jun 1998.

\bibitem{hht}''\textbf{Quantum Computation in Brain Microtubules? Decoherence
and Biological Feasibility''} by \emph{S. Hagan, S. R. Hameroff,
}and\emph{\ J. A. Tuszy\'{n}ski}, LANL e-print archives; quant-ph/0005025, 4
May 2000.

\bibitem{zizzi}''\textbf{Holography, Quantum Geometry, and Quantum Information
Theory''} by \emph{Paola A. Zizzi}, LANL e-print archives; arXiv:
gr-qc/9907063 v2, 31 Mar 2000.

\bibitem{zizzi2}''\textbf{Emergent Consciousness: From the Early Universe to
Our Mind''} by \emph{Paola A. Zizzi}, LANL e-print archives; arXiv:
gr-qc/0007006, 7 Jul 2000.

\bibitem{benni}''\textbf{Distillation of vacuum entanglement to EPR pairs''}
by \emph{Benni Reznick}, LANL e-print archives; quant-ph/0008006, 1 Aug 2000.
\end{thebibliography}
\end{document}